\PassOptionsToPackage{dvipsnames}{xcolor}
\documentclass[a4paper,11pt]{article}
\usepackage{amsmath,amssymb,amsfonts,amsthm,latexsym} 

\usepackage{pos}
\usepackage[normalem]{ulem}
\usepackage[utf8]{inputenc}
\usepackage[T1]{fontenc}
\usepackage{cancel}
\usepackage{graphicx}
\usepackage{latexsym}
\usepackage{bbm}
\usepackage{bm}
\usepackage{epsfig}
\usepackage{epstopdf}
\usepackage{subcaption}
\usepackage{color}
\usepackage{comment}
\usepackage{slashed}
\usepackage{placeins}
\usepackage{cleveref}
\usepackage{setspace}
\usepackage{url}
\usepackage{tabularx}
\usepackage{enumitem}
\usepackage{hyperref}
\usepackage{diagbox}
\usepackage{dcolumn}
\usepackage{xfrac}
\usepackage{pgf}

\newcommand{\tr}{\mathop{\mathrm{tr}}}

\newcommand{\Nf}{N_{\rm f}}
\newcommand{\Nc}{N_{\rm c}}
\newcommand{\Ns}{N_{\rm s}}
\newcommand{\Nt}{N_{\rm t}}
\newcommand{\CA}{C_{\rm A}}

\newcommand{\CF}{C_{\rm F}}

\newcommand{\e}[1]{\text{e}^{#1}}

\newcommand{\as}{\alpha_{\mathrm{s}}}
\newcommand{\als}{\alpha_{\mathrm{s}}}

\newcommand{\MSb}{\overline{\textrm{MS}}}
\newcommand{\QCD}{{\MSb}}

\newcommand{\be}{\begin{equation}} 
\newcommand{\ee}{\end{equation}}
\def\ba#1\ea{\begin{align}#1\end{align}}
\newcommand{\bea}{\begin{eqnarray}} 
\newcommand{\eea}{\end{eqnarray}}
\def\lsim{\mathrel{\raise.3ex\hbox{$<$\kern-.75em\lower1ex\hbox{$\sim$}}}}
\def\gsim{\mathrel{\raise.3ex\hbox{$>$\kern-.75em\lower1ex\hbox{$\sim$}}}}

\renewcommand{\sfrac}[2]{\ensuremath{#1/#2}}

\renewcommand{\sfrac}[2]{\ensuremath{#1/#2}}
\newcommand{\Nstates}{\ensuremath{N_{\text{st}}}}

\colorlet{preliminary}{White!92!BrickRed}

\newcommand{\st}[1]{}

\let\oldcite\cite
\renewcommand{\cite}[1]{\mbox{\oldcite{#1}}}

\title{Strong coupling\st{ constant} 
from the one-loop improved static energy}
\author*[a,b]{Viljami~Leino}
\author[c]{Alexei~Bazavov}
\author[d,e,f]{Nora~Brambilla}
\author[g]{Georg~von~Hippel}
\author[h,f]{Andreas~S.~Kronfeld}
\author[d,e]{Julian~Mayer-Steudte}
\author[i]{Peter~Petreczky}
\author[d]{Sipaz~Sharma}
\author[d,j]{Sebastian~Steinbeißer}
\author[d]{Antonio~Vairo}
\author[k,l]{Johannes~H.~Weber}

\affiliation[a]{Quantum Theory Center ($\hbar$QTC), University of Southern Denmark, 5230 Odense M, Denmark}
\affiliation[b]{Dept. of Mathematics and Computer Science, University of Southern Denmark, 5230 Odense M, Denmark
}
\affiliation[c]{Department of Computational Mathematics, Science and Engineering, and Department of \\
Physics and Astronomy, Michigan State University, East Lansing, Michigan 48824, USA} 
\affiliation[d]{Physics Department, TUM School of Natural Sciences, 
Technical University of Munich, 
\\ 
James-Franck-Straße~1, 85748 Garching b.\ München, Germany}
\affiliation[e]{Munich Data Science Institute, Technical University of Munich, \\ 
Walther-von-Dyck-Straße~10, 85748 Garching b.\ München, Germany}
\affiliation[f]{Institute for Advanced Study, Technical University of Munich, \\ 
Lichtenbergstraße~2a, 85748 Garching b.\ München, Germany}
\affiliation[g]{PRISMA+ Cluster of Excellence and Institut für Kernphysik, Johannes Gutenberg-Universität Mainz,\\
             Johann-Joachim-Becher-Weg 48, 55128 Mainz, Germany}
\affiliation[h]{Particle Theory Department, Theory Division, Fermi National Accelerator Laboratory, \\
    Batavia, Illinois 60510-5011, USA}
\affiliation[i]{Physics Department, Brookhaven National Laboratory, Upton, New York 11973-5000, USA}
\affiliation[j]{Leibniz-Rechenzentrum der Bayerischen Akademie der Wissenschaften, \\ 
Boltzmannstraße~1, 85748 Garching b.\ München, Germany}
\affiliation[k]{Institut f\"ur Kernphysik, Technische Universit\"at Darmstadt, Schlossgartenstra\ss e 2, 64289 Darmstadt, Germany}
\affiliation[l]{Werner-Heisenberg-Gymnasium Bad D{\"u}rkheim, Kanalstra{\ss}e 19, 67098 Bad D{\"u}rkheim, Germany}

\emailAdd{leino@qtc.sdu.dk}

\abstract{
\vspace*{-1mm}
\textbf{\textsf{TUMQCD Collaboration}}\\[0.25em]
The static energy is an excellent observable for extracting the strong coupling $\as$ on the lattice. 
For short distances, the static energy can be calculated both on the lattice using Wilson line correlators, 
and with perturbation theory up to three-loop accuracy with leading ultrasoft log resummation. 
Comparing the perturbative expression and lattice data allows for precise determination of $\as$. 
We present preliminary results for one-loop lattice perturbation theory improvement of the Wilson loop
and show how it improves the $\as$ extraction. 
We present a preliminary reanalysis of the TUMQCD (2+1)-flavor QCD data.}

\FullConference{The 42nd International Symposium on Lattice Field Theory (LATTICE2025)\\
2-8 November 2025\\
Tata Institute of Fundamental Research, Mumbai, India\\}

\begin{document}

\maketitle

\section{Introduction}
The static energy $E_0(r)$ between a static quark and a static antiquark separated by a distance $r$ is a fundamental 
observable that has played an important role~\cite{Bali:2000gf} in establishing confinement in QCD.
At short distances the static energy is well defined in perturbative weak coupling expansion up to $\mathrm{N}^3\mathrm{LL}$ level~\cite{Brambilla:1999qa,Pineda:2000gza,Brambilla:2006wp,Brambilla:2009bi,Anzai:2009tm,Smirnov:2009fh}. 
On the lattice, $E_0(r)$ can be measured with high precision as the ground state of a static Wilson loop. 
As the static energy is a scheme invariant observable, we can measure the strong coupling\st{ constant} $\as$ by comparing
the lattice measurements with perturbation theory in the range of separations where both methods are reliable.

The strong coupling $\as$ is a fundamental parameter of QCD and the Standard Model of particle physics.
The running of the strong coupling in the $\MSb$ scheme is a function of the renormalization scale $\mu$ and the intrinsic scale of QCD,
$\Lambda_{\QCD}$.
In this proceedings, we focus on determining this intrinsic scale from the static potential. 
When $\Lambda_{\QCD}$ is known, one can perturbatively determine $\as$ at any scale $\mu\gg\Lambda_{\QCD}$.
For reviews of the current status of $\as$ from the lattice and experiments, we refer the reader to the review 
articles~\cite{Aoki:2024oxs,dEnterria:2022hzv}.

So far, $\Lambda_{\MSb}$ has been determined from the static energy in $\Nf=0$ pure gauge SU(3) Yang-Mills 
theory~\cite{Brambilla:2010pp,Husung:2017qjz,Brambilla:2023fsi} and with either 
$\Nf=2$ dynamical quark flavors~\cite{Jansen:2011vv,Karbstein:2014bsa,Karbstein:2018mzo} or with 
$\Nf=2+1$ dynamical flavors~\cite{Bazavov:2012ka,Bazavov:2014soa,Takaura:2018vcy,Bazavov:2019qoo,Ayala:2020odx,Mena-Valle:2025hky}.
A very preliminary determination by TUMQCD at $\Nf=2+1+1$ was presented in recent proceedings~\cite{Leino:2025pvl}.

The extraction of $\as$ from the static energy requires data at very small distances. 
However, the rotational symmetry is broken on the lattice and the static energy at small distances $r$ exhibits significant non-smooth discretization effects. These cutoff effects are known to leading order and can be removed at tree-level by defining an improved distance $r_I$ such that the continuum theory matches the lattice theory at leading order. 
For the small distances required for $\as$ extraction, the tree-level improvement is not enough~\cite{Bazavov:2019qoo}.
We compute the one-loop improvement of the static Wilson line correlators for our ensembles to reduce the discretization errors at short distances.
The Wilson loop has been calculated at one-loop level before, 
for pure gauge theories~\cite{Curci:1983an,Heller:1984hx,Weisz:1983bn,Snippe:1997ru},
and for the Wilson quarks~\cite{Martinelli:1998vt},
overlap quarks~\cite{Athenodorou:2005hi}, and the regular staggered quarks~\cite{Bali:2002wf}. 
However, the one-loop improvement is not known for the commonly used highly improved staggered quark (HISQ)~\cite{Follana:2006rc} ensembles. In these proceedings we show the preliminary result for the one-loop improvement of Wilson line correlators for HISQ ensembles~\cite{lpt_paper}. The improvement is then applied for a preliminary extraction of $\Lambda_{\QCD}$.

\section{Static energy}
\subsection{Perturbation theory}
The static energy $E_0(r)$ has the perturbative expansion
\begin{align}\label{eq:statenergyI}
E_0(\bm{r}) = \Lambda_0 - \frac{\CF}{r} \left[ 1 + \frac{a^2}{r^2}
\sum_{k=0} \frac{g_X^{2k}}{(4\pi)^k} \hat{x}_k(\bm{r}\st{r}/a)\right]\left(\sum_{l=0} \frac{g_X^{2l+2}(\mu)}{(4\pi)^{l}} v_{X,l}(r/a)\right)\,.
\end{align}
where $\Lambda_0$ is a constant of mass dimension one and the coefficients, for scheme $X$, $v_{X,l}$ can be found for the $\MSb$ scheme in the literature~\cite{Brambilla:1999qa,Pineda:2000gza,Brambilla:2006wp,Brambilla:2009bi,Anzai:2009tm,Smirnov:2009fh}.
Starting at three loops, the perturbation theory will have ultrasoft logarithms $\ln[\frac{1}{2}\CA \als(\mu_\mathrm{us})]$.
We set the soft scale to be $\mu=1/r$ and the ultrasoft scale $\mu_\mathrm{us}=\CA\as(\mu)/2r$.  
The coefficients $\hat{x}_k$ describe the rotation symmetry breaking discretization effects arising from lattice perturbation theory.
In rotationally symmetric schemes all $\hat{x}_k=0$.

The constant $\Lambda$ in Eq.~\eqref{eq:statenergyI} is a scheme dependent quantity. 
In dimensional regularization it manifests as a renormalon of mass dimension one and
on the lattice it is related to a linear divergence in the inverse lattice spacing $a^{-1}$.
Since all the $\as$ dependence is contained in the $r$-dependent part of $E_0(r)$, 
we can obtain $\as$ from the lattice data by  matching the lattice results and perturbative curves at some reference distance $r^\ast$.
However, in order to get as stable as possible perturbative behavior, the issue with the renormalon has to be addressed. 
We explore two strategies to deal with the renormalon contribution:
\begin{enumerate}
\item By taking a derivative with respect to $r$ of $E_0(r)$, we define the static force $F(r)$.
The derivative naturally removes the constant $\Lambda_0$. While $F(r)$ can be computed directly on the
lattice~\cite{Brambilla:2021wqs,Brambilla:2023fsi}, in these proceedings we focus on analyzing the static energy and
we integrate numerically the perturbative expression with running $\as$ of the static force to get a more stable definition of the static energy. 
While the constant reappears, now as an integration constant, this expression has reduced renormalon contributions. 
For more details, see Refs.~\cite{Bazavov:2014soa,Bazavov:2019qoo}.
\item Alternatively, we can apply the minimal renormalon subtraction (MRS) prescription~\cite{Brambilla:2017hcq,Komijani:2017vep,Kronfeld:2023jab}. 
This method also derives the static force but integrates in an analytic way that exposes the leading factorial growth of the expansion coefficients. 
The factorial growth is summed to all orders, which yields an unambiguous definition of $\Lambda$ 
and reduces the perturbative error of the fits. 
\end{enumerate}
\begin{figure}
    \centering
    \includegraphics[width=0.49\textwidth]{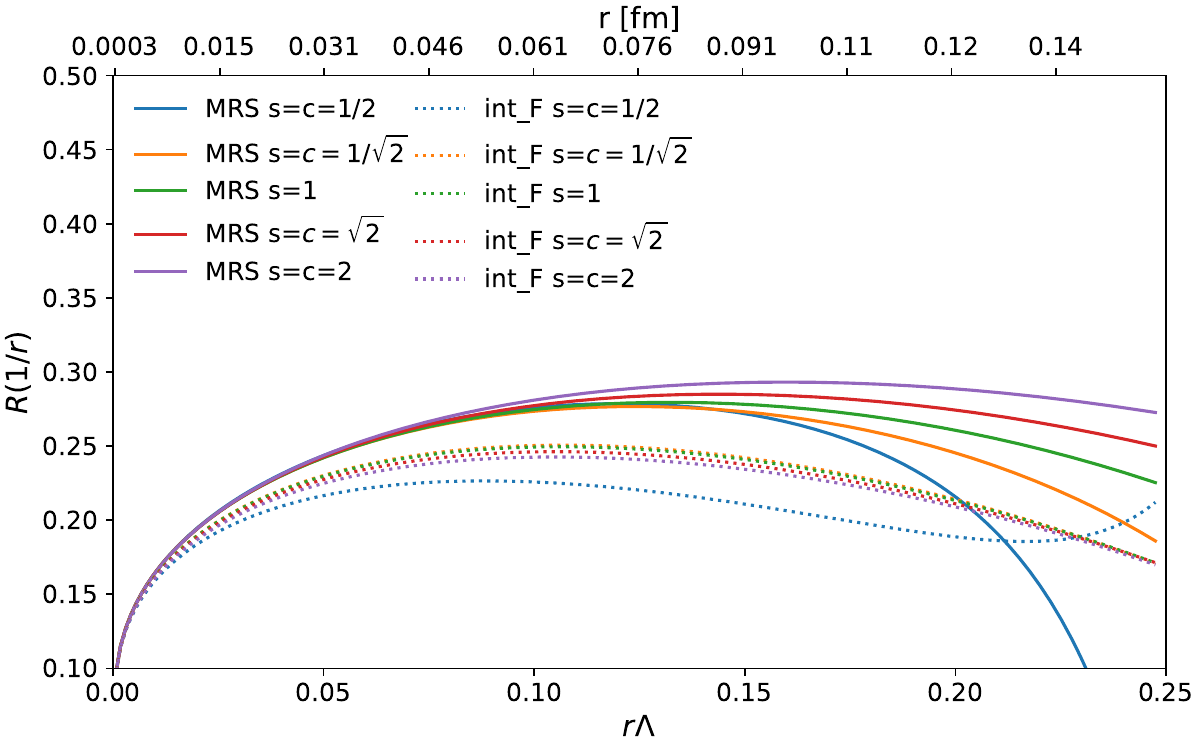}
    \includegraphics[width=0.49\textwidth]{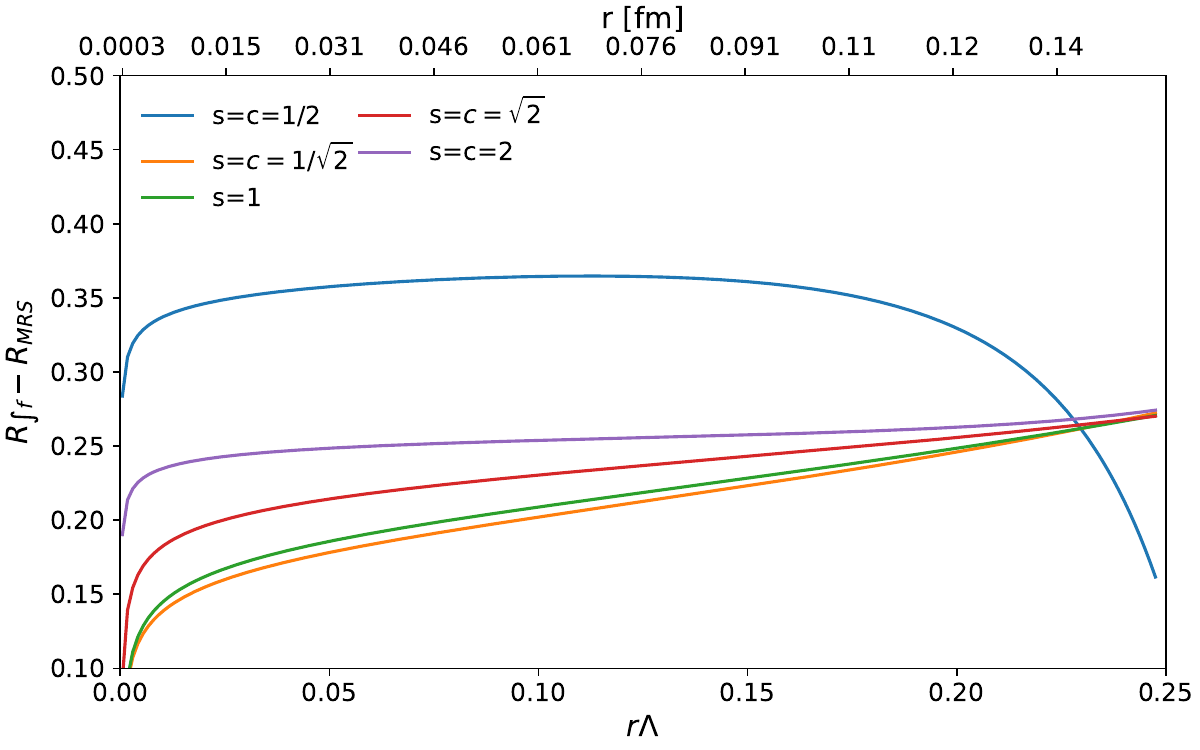}
    \caption{Left: The leading factor normalized static energy $R=-rE_0/\CF$ for the two different approaches to reduce the renormalon contribution. Right: The difference between the two methods.}
    \label{fig:pert}
\end{figure}
With these procedures the scale variation dependence of the perturbative series should be reduced.
We present this scale dependence in Fig.~\ref{fig:pert}. On the left, we show both the MRS and the integrated force at different scales $\mu=s/r$. When extracting $\Lambda_{\MSb}$, we will need to choose appropriate distance $r\sim0.13$~fm where we have enough points for a stable fit and still small enough systematic error from the scale variance.  
While both of these methods eliminate the renormalon ambiguity, they still retain an arbitrary constant. It is this remaining constant that explains the slight difference between $F(r)$ and MRS on the left side of the figure. When comparing to lattice data, the data needs to be shifted to either of these curve clusters. To understand the differences between these methods better, we show on the right side of the figure, the difference between the two methods. While the difference is nearly constant at small separations, we observe MRS to be slightly more stable towards the larger distances. The MRS method is also numerically faster to calculate than numerically integrating the force. Therefore, we will use the MRS approach for the rest of the proceedings.

\subsection{One-loop lattice perturbation theory}
We calculate the one-loop improvement numerically using the program \texttt{HPsrc}~\cite{Hart:2009nr}. 
Due to the complexity of the Feynman rules for the HISQ action, we employ \texttt{HiPPy}~\cite{Hart:2009nr} 
to automatically derive them.
To take finite volumes into account, we calculate the momentum exchange between the static lines at discrete values, while the loop momenta are integrated at continuum with \texttt{VEGAS}. The only exception to this are the fermion loops in the gluon polarization which are unstable at very small momenta.
For the fermionic gluon polarization we therefore also calculate discrete sums for the loop integrals and use their infinite volume extrapolation at small loop momenta. We check in the intermediate region of momenta that the discrete sums and \texttt{VEGAS} results agree.
After we have measured all the diagrams in the momentum space, we take a fast Fourier transform (FFT) to transform to position space static energy. As we cannot measure the momentum zero modes, we set
them to zero. This causes the final position space static energy to have an arbitrary constant contribution. Since the real lattice data and the continuum perturbation theory also have arbitrary constants, we can ignore the constant for now and fit it away later by matching to lattice data.
After FFT, we either interpolate to the actual lattice volume of a particular ensemble 
or extrapolate to infinite volume with a linear ansatz in the inverse volume. For these proceedings we show results for the infinite volume extrapolated improvement. 

\begin{figure}
    \centering
    \includegraphics[width=0.49\textwidth]{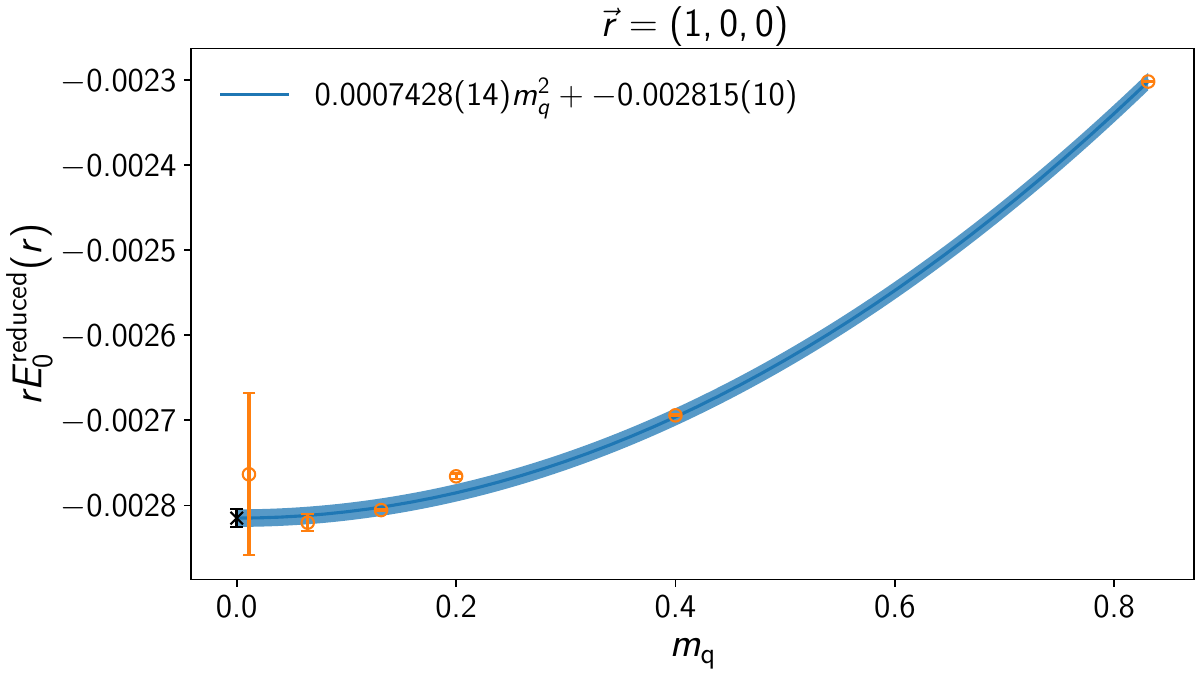}
    \includegraphics[width=0.49\textwidth]{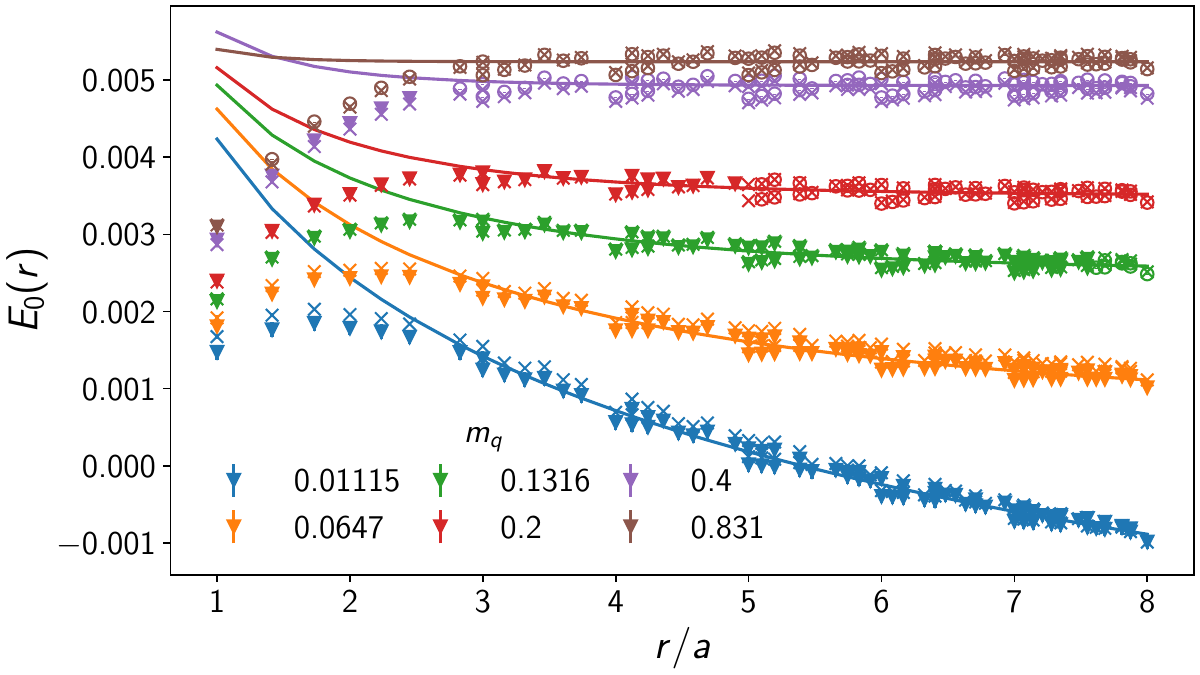}
    \vspace{-2.1cm}
    \begin{center}
    \hspace{3cm}\textcolor{preliminary}{\rotatebox{15}{PRELIMINARY}}\hspace{6cm}\textcolor{preliminary}{\rotatebox{20}{PRELIMINARY}}
    \end{center}
    \vspace{0.0cm}
    \caption{Left: the interpolation of residual finite mass effects for 1-loop lattice perturbation theory. Right: The final interpolation reconstruction of the fermionic contribution (x's) compared to data from \texttt{HPsrc} (solid symbols). The solid line shows the perturbative curve that we are expected to approach asymptotically.}
    \label{fig:lptf}
\end{figure}

We measure the set of fermions individually in a finite set of seven masses spanning from the smallest to the largest mass in the ensembles of interest. Since the set of available ensembles has more masses than the seven simulated ones, we need to interpolate to an arbitrary mass. Moreover, we note that the simulations at nearly zero mass are very noisy, but the respective mass contribution is small; hence we combine the two smallest masses as an estimate of zero mass contribution. The fit procedure goes as follows. First, we fit the generic logarithmic running for a fermion with a specific mass $am_q$
\begin{equation}\label{eq:Elogfit}
rE_{0}^\text{log-fit}(m_q,r) = -\frac{2}{(4\pi)^3}\CF
\beta_0^\mathrm{f} \left( \delta_{E_{0}}(m_q,r) + A\right) +Cr \,,
\end{equation}
where $A$ is the main fit parameter, $C$ accounts for the arbitrary shift, and everything else is determined by perturbation theory.
The one-loop finite mass correction term $\delta_{E_0}(m_q,r)$, which interpolates between the logarithmic running of zero mass fermion and the constant zero for the completely decoupled fermion, is taken from the literature~\cite{Eiras:1999xx}. The superscript f in perturbative coefficients indicate that only the $\Nf$ dependent part of the coefficient is taken into account. We show some of our reference masses on the right side of Fig.~\ref{fig:lptf}, where the filled symbols present the data from \texttt{HPsrc} and the solid lines show the fit~\eqref{eq:Elogfit}. Since the constant $C$ is arbitrary and will be matched again at the later stages of the analysis, we have subtracted it from the data in order to establish a common scheme in terms of the constant contribution. 
The fit parameter $A$ is then interpolated to an arbitrary mass as $A=\log(A_2 m_q^2 + A_1 m_q + A_0)$ with $A_i$ being fit parameters. 

Next, we note that most of mass dependence is already captured by the perturbative curve and we define a reduced quantity 
and then interpolate the remaining mass dependence
\begin{align}
E_{0}^\mathrm{reduced}(m_q,r) &= E_{0}(m,r) - E_{0}^\text{log-fit}(m,r) \,, \\
rE_{0}^\mathrm{reduced}(m_q,r) &= B(r) m_q^2 + D(r)\,.
\end{align}
On the left side of  Fig.~\ref{fig:lptf}, we show this fit for the shortest distance $r/a=1$, which is the distance where this effect is the largest. As distance increases, the $m_q^2$ term decreases.
We see that the larger masses regulate the noisy zero mass contribution and that below $am_q <0.1$ the reduced mass contribution is nearly constant. 
Lastly, on the right side of Fig.~\ref{fig:lptf}, we reconstruct the original \texttt{HPsrc} data from purely interpolated
numbers (shown as x's) to show the interpolation process works. We especially see\st{,} that despite combining the two smallest masses to estimate zero mass, we still perfectly reconstruct the $am_q =0.01115$ behavior.
\begin{figure}
    \centering
    \includegraphics[width=0.49\textwidth]{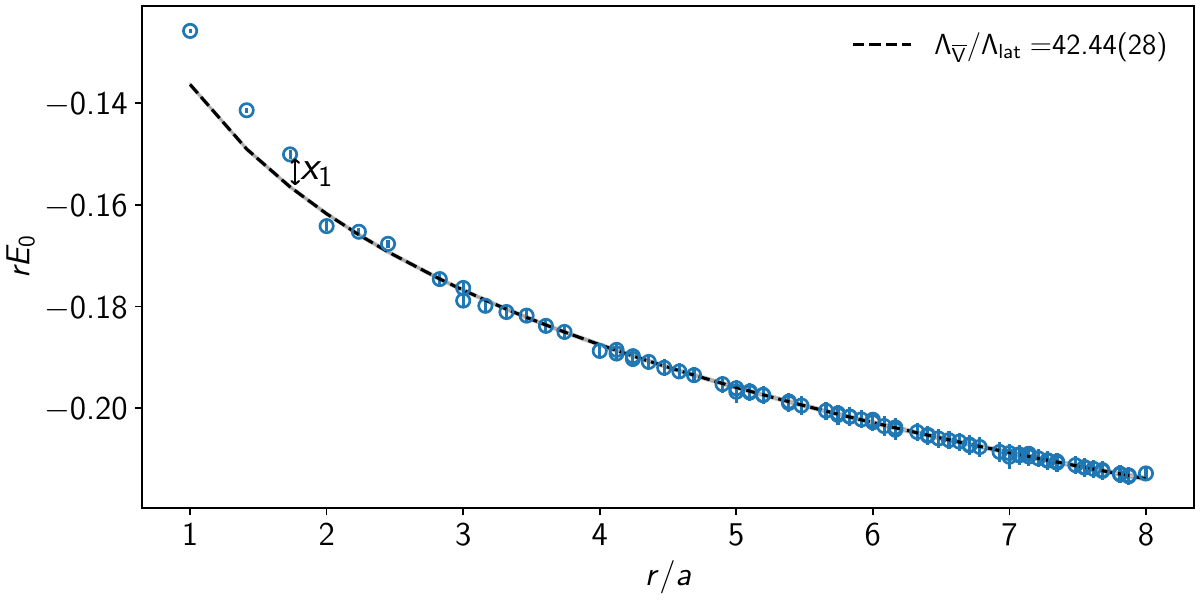}
    \includegraphics[width=0.49\textwidth]{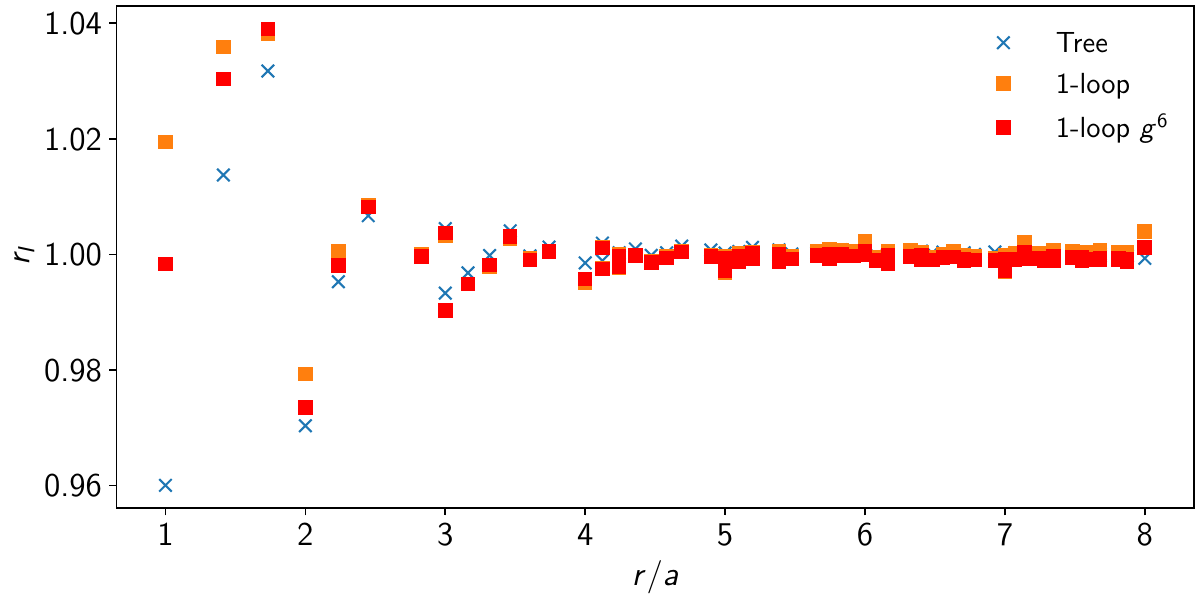}
    \vspace{-2.9cm}
    \begin{center}
    \hspace{3cm}\textcolor{preliminary}{\rotatebox{15}{PRELIMINARY}}\hspace{4cm}\textcolor{preliminary}{\rotatebox{-10}{PRELIMINARY}}
    \end{center}
    \vspace{0.6cm}
    \caption{Left: Example of the final 1-loop lattice perturbation theory contribution for a single ensemble and a fit to  $\Lambda$-ratio. Right: The final 1-loop improvement compared to the tree-level improvement.}
    \label{fig:lpt}
\end{figure}

After the fermions have been interpolated, we can create the one-loop lattice perturbation theory contribution to an arbitrary ensemble by combining the right amount of quarks with their respective masses with the gluonic contribution. We show this data on the left side of Fig.~\ref{fig:lpt}. This data should asymptotically converge towards the continuum perturbation theory. Therefore, we fit\st{:}
\begin{equation}
rE_{0}^\text{1-loop} = -\frac{2\CF\beta_0}{(4\pi)^3}\left[ \ln \frac{\Lambda_{\overline{V}}}{\Lambda_\mathrm{lat}} + \ln \frac{r}{a}\right]
+ rC = \frac{4\pi}{\CF} v_{\mathrm{lat},1} \,,
\end{equation}
for the ratio of $\Lambda$-parameters between the $\overline{V}$-scheme 
and the lattice scheme and the arbitrary shift $C$.
The logarithm $\ln(r/a)$ portrays the fact that $\overline{V}$-scheme runs with $\mu=1/r$, while the lattice scheme naturally abides to scale $\mu=1/a$. We show this fit in Fig.~\ref{fig:lpt},
where we can observe that indeed the asymptotic is well matched.
We can then define the one-loop discretization effect coefficient in Eq.~\eqref{eq:statenergyI} $\hat{x}_1 = -4\pi r/\CF (E_{0,\mathrm{lat}}^\text{1-loop}(r/a) - v_\mathrm{lat,1}E_{0,\mathrm{lat}}^\mathrm{tree})$. Combining with known tree-level improvement $\hat{x}_0$, we can define improved distance $r_\mathrm{I}^\text{1-loop}$ such that it makes the square bracket of Eq.~\eqref{eq:statenergyI} equal to unity. 
Alternatively, we can first calculate the product of the sums in Eq.~\eqref{eq:statenergyI} and define the improved distance such that all terms with $\hat{x}_k$ vanish. These two approaches differ by $g^6$ terms, which are beyond the accuracy of the one-loop improvement. 
We show the resulting improved distances on the right side of Fig.~\ref{fig:lpt}. We observe that at large distances, the tree level improvement is enough but at small distances the one-loop improvement has sizable effect. It is possible that at the very short distances $r\approx a$, the one-loop improvement is not quite enough and higher perturbative orders would be necessary as indicated by the size of the $g^6$ effects.
To account for these higher order effects, for the later analysis shown in these proceedings, we boost our short distance $r^2\le 6a$ lattice data errors by $0.5\%$. 

\subsection{Lattice data}
We compute the static energy from the (2+1)-flavor MILC and HotQCD lattice ensembles that we used previously 
in Refs.~\cite{Bazavov:2014soa,Bazavov:2019qoo}. 
For gluons the action is Symanzik-improved action, while the sea quarks are simulated with  
the HISQ-action~\cite{Follana:2006rc}. 
The gauge configurations have been fixed to Coulomb gauge, which allows easy access to off-axis distances.
Instead of  Wilson loops, in Coulomb gauge, $E_0(r)$ can be obtained from the time dependence of the Wilson-line correlation function
$C\left(\bm{r},\tau,a\right)$ at separation $\sfrac{\bm{r}}{a}$:
\begin{align}
    C\left(\bm{r},\tau,a\right) &= \left\langle \frac{1}{\Ns^{3}} \sum\limits_{\bm{x}} \sum\limits_{\bm{y}=R(\bm{r})}
        \frac{1}{\Nc\,N_{\bm{r}}} \tr\left[W^{\dagger}\left(\bm{x}+\bm{y},\tau,a\right)
            W\left(\bm{x},\tau,a\right)\right] \right\rangle
    \nonumber \\
        &= \sum_{n=0}^{\infty} C_{n}\left(\bm{r},a\right) \left(\e{-\tau E_{n}\left(\bm{r},a\right)} +
            \e{-(a\Nt-\tau)E_{n}\left(\bm{r},a\right)} \right)
    \nonumber \\
     &= \e{-\tau E_{0}\left(\bm{r},a\right)} \left( C_{0}\left(\bm{r},a\right) +
        \sum\limits_{n=1}^{\Nstates-1} C_{n}\left(\bm{r},a\right)
            \prod\limits_{m=1}^{n} \e{-\tau \Delta_{m}\left(\bm{r},a\right)}
        \right) + \ldots,
    \label{eq:forward_tower}
\end{align}
where we have re-parameterized the correlation in terms of energy differences $a\Delta_{n}(\bm{r},a)=aE_{n}(\bm{r},a)-aE_{(n-1)}(\bm{r},a)>0$.
The reader is referred to earlier TUMQCD publications~\cite{Bazavov:2014soa,Bazavov:2019qoo} for further details.
For these proceedings, we focus on showing fits done self-consistently with $r_1$-units. The physical value of 
the $r_1$ with $2+1$~fermions\st{ case} has been recently updated~\cite{Larsen:2025wvg}.

\section{Extraction of \texorpdfstring{\boldmath$\Lambda_{\MSb}$}{Lambda MSbar}}
To extract $\Lambda_{\MSb}$, we perform a joint fit to all ensembles.
We fit a shared $\Lambda_{\MSb}$ 
for all ensembles. Furthermore, we also include arbitrary shift parameter, that matches the lattice data to perturbative curve, for each ensemble. We evaluate the static potential at three-loop order.
For the fits shown in the proceedings, we do not include any further terms to describe lattice artifacts beyond the tree-level and one-loop improvement.

To keep the scale dependence down and to stay in the perturbative range,
we limit the fits to maximum separation of $r<0.15$~fm and fit over all possible data ranges. 
The fits from different ranges are then combined with a model average using the Akaike information criterion (AIC) 
to weight the different fits~\cite{Jay:2020jkz}. 
\begin{figure}
    \centering
    \includegraphics[width=0.5\textwidth]{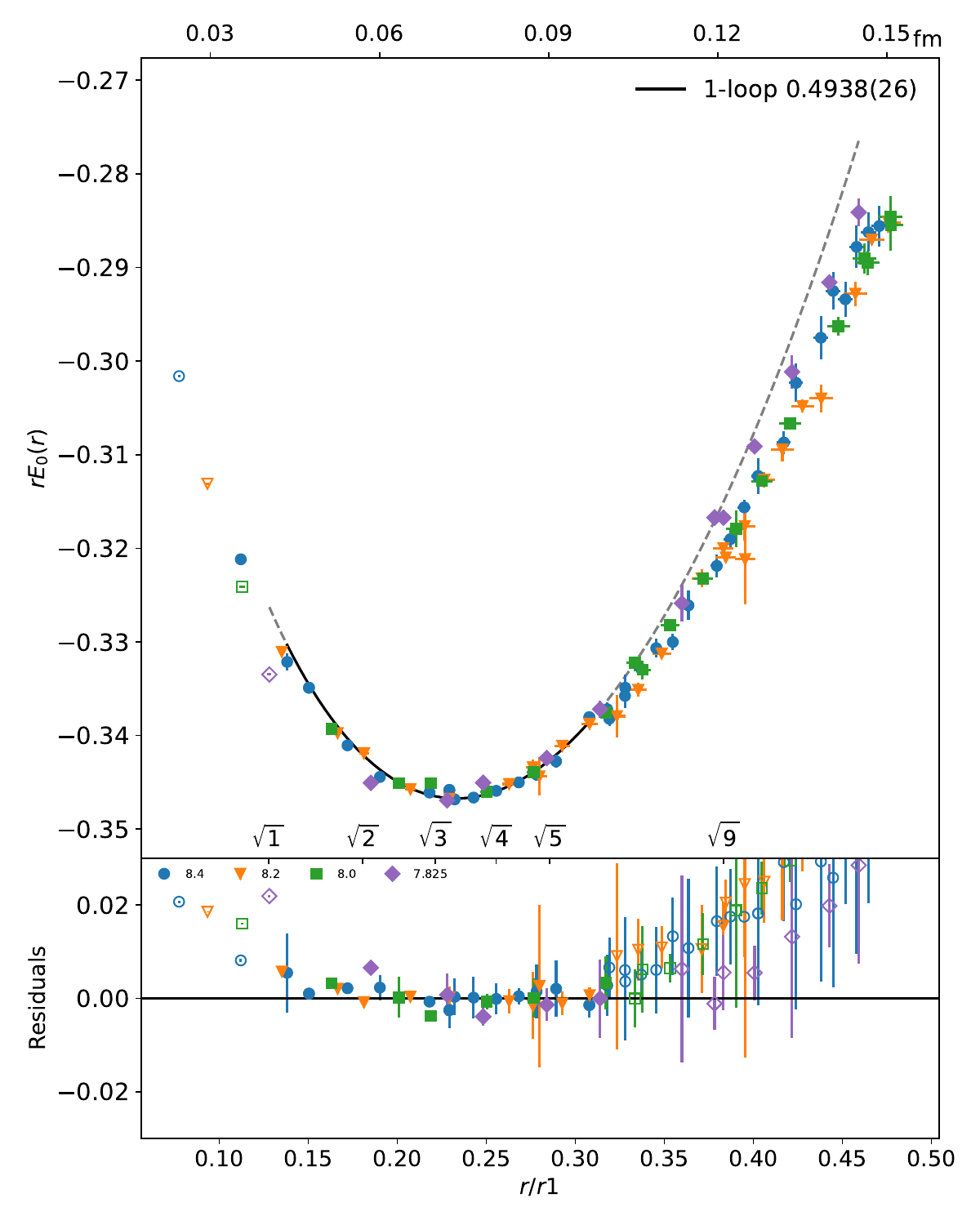}
    \vspace{-7cm}
    \begin{center}
    \hspace{-0cm}\textcolor{preliminary}{\rotatebox{-30}{PRELIMINARY}}
    \end{center}
    \vspace{4.5cm}
    \caption{Example of the $\Lambda$ extraction from a joint fit over 4 finest ensembles with 1-loop improved distances.}
    \label{fig:lambdafit}
\end{figure}
This fit procedure is demonstrated in Fig.~\ref{fig:lambdafit}.
The colored points show the lattice data for each ensemble, with the filled points being included in the fit and the empty points at separation $r=a$ being excluded due to unknown higher order discretization effects. On the lower row of the figure,
the filled points present the data included in the fit indicated by the solid line and the empty symbols match the regimes shown by the dashed lines. 
The solid curve shows the range of the fit with best AIC weight and the dashed lines are there just to guide the reader's eye.
The resulting $r_1 \Lambda\approx 0.494(3)$ is in good agreement with our previous extractions~\cite{Bazavov:2012ka,Bazavov:2014soa,Bazavov:2019qoo}. We note that this is an individual fit with no analysis on the  systematic effects and expect the final error with systematics included to be somewhat larger than what is shown in the plot. The main takeaway of these proceedings is to showcase the benefits of one-loop improved distance in the $\as$ extraction.

\acknowledgments{
The lattice QCD calculations have been performed using the publicly available \href{https://web.physics.utah.edu/~detar/milc/milcv7.html}{MILC code}. 
The simulations were carried out on the computing facilities of the Computational Center for Particle and Astrophysics (C2PAP) in the project 
\emph{Calculation of finite $T$ QCD correlators} (pr83pu) and of the SuperMUC cluster at the Leibniz-Rechenzentrum (LRZ) in the project
\emph{The role of the charm-quark for the QCD coupling constant} (pn56bo),
both located in Munich (Germany),
The authors acknowledge the Gauss Centre for Supercomputing e.V.
(\href{www.gauss-centre.eu}{www.gauss-centre.eu})
for funding this project by providing computing time on the GCS Supercomputer SuperMUC-NG
at Leibniz Supercomputing Centre (\href{www.lrz.de}{www.lrz.de}).
This research was funded by the Deutsche Forschungsgemeinschaft (DFG, German Research Foundation) cluster of excellence "ORIGINS" (\href{https://www.origins-cluster.de}{www.origins-cluster.de}) under Germany's Excellence Strategy EXC-2094-390783311.
N. B. acknowledges the European Research Council advanced grant~ERC-2023-ADG-Project~EFT-XYZ.
P.P. was supported by The U.S. Department of Energy through Contract No.~DE-SC0012704.
A.~B.'s work was supported by the US National Science Foundation under grant No.~PHY23-09946.
This document was prepared by TUMQCD using the resources of the Fermi National Accelerator Laboratory (Fermilab), a U.S. Department of Energy, Office of Science, Office of High Energy Physics HEP User Facility. Fermilab is managed by Fermi Forward Discovery Group, LLC, acting under Contract No.~89243024CSC000002. The work of V.L. is supported by the Carlsberg Foundation grant CF22-0922.


\bibliographystyle{jhep_modified}
\bibliography{alphas.bib}

\providecommand{\href}[2]{#2}\begingroup\raggedright\begin{thebibliography}{10}

\bibitem{Bali:2000gf}
G.~S. Bali, \emph{{QCD} forces and heavy quark bound states},
  \href{https://doi.org/10.1016/S0370-1573(00)00079-X}{\emph{Phys. Rept.}
  {\bfseries 343} (2001) 1}
  [\href{https://arxiv.org/abs/hep-ph/0001312}{{\ttfamily hep-ph/0001312}}].
\bibitem{Brambilla:1999qa}
N.~Brambilla, A.~Pineda, J.~Soto and A.~Vairo, \emph{The infrared behavior of
  the static potential in perturbative {QCD}},
  \href{https://doi.org/10.1103/PhysRevD.60.091502}{\emph{Phys. Rev. D}
  {\bfseries 60} (1999) 091502}
  [\href{https://arxiv.org/abs/hep-ph/9903355}{{\ttfamily hep-ph/9903355}}].
\bibitem{Pineda:2000gza}
A.~Pineda and J.~Soto, \emph{The renormalization group improvement of the {QCD}
  static potentials},
  \href{https://doi.org/10.1016/S0370-2693(00)01261-2}{\emph{Phys. Lett. B}
  {\bfseries 495} (2000) 323}
  [\href{https://arxiv.org/abs/hep-ph/0007197}{{\ttfamily hep-ph/0007197}}].
\bibitem{Brambilla:2006wp}
N.~Brambilla, X.~Garcia~i Tormo, J.~Soto and A.~Vairo, \emph{The logarithmic
  contribution to the {QCD} static energy at {N$^{4}$LO}},
  \href{https://doi.org/10.1016/j.physletb.2007.02.015}{\emph{Phys. Lett. B}
  {\bfseries 647} (2007) 185}
  [\href{https://arxiv.org/abs/hep-ph/0610143}{{\ttfamily hep-ph/0610143}}].
\bibitem{Brambilla:2009bi}
N.~Brambilla, A.~Vairo, X.~Garcia~i Tormo and J.~Soto, \emph{The {QCD} static
  energy at {N$^{3}$LL}},
  \href{https://doi.org/10.1103/PhysRevD.80.034016}{\emph{Phys. Rev. D}
  {\bfseries 80} (2009) 034016}
  [\href{https://arxiv.org/abs/0906.1390}{{\ttfamily 0906.1390}}].
\bibitem{Anzai:2009tm}
C.~Anzai, Y.~Kiyo and Y.~Sumino, \emph{Static {QCD} potential at three-loop
  order}, \href{https://doi.org/10.1103/PhysRevLett.104.112003}{\emph{Phys.
  Rev. Lett.} {\bfseries 104} (2010) 112003}
  [\href{https://arxiv.org/abs/0911.4335}{{\ttfamily 0911.4335}}].
\bibitem{Smirnov:2009fh}
A.~V. Smirnov, V.~A. Smirnov and M.~Steinhauser, \emph{Three-loop static
  potential}, \href{https://doi.org/10.1103/PhysRevLett.104.112002}{\emph{Phys.
  Rev. Lett.} {\bfseries 104} (2010) 112002}
  [\href{https://arxiv.org/abs/0911.4742}{{\ttfamily 0911.4742}}].
\bibitem{Aoki:2024oxs}
{\scshape Flavour Lattice Averaging Group (FLAG)} collaboration, Y.~Aoki
  et~al., \emph{{FLAG Review 2024}},
  \href{https://arxiv.org/abs/2411.04268}{{\ttfamily 2411.04268}}.
\bibitem{dEnterria:2022hzv}
D.~d'Enterria et~al., \emph{{The strong coupling constant: State of the art and
  the decade ahead}}, \href{https://doi.org/10.1088/1361-6471/ad1a78}{\emph{J.
  Phys. G} {\bfseries 51} (2024) 090501}
  [\href{https://arxiv.org/abs/2203.08271}{{\ttfamily 2203.08271}}].
\bibitem{Brambilla:2010pp}
N.~Brambilla, X.~Garcia~i Tormo, J.~Soto and A.~Vairo, \emph{{Precision
  determination of $r_0\Lambda_{\overline{\rm MS}}$ from the QCD static
  energy}}, \href{https://doi.org/10.1103/PhysRevLett.105.212001}{\emph{Phys.
  Rev. Lett.} {\bfseries 105} (2010) 212001}
  [\href{https://arxiv.org/abs/1006.2066}{{\ttfamily 1006.2066}}].
\bibitem{Husung:2017qjz}
N.~Husung, M.~Koren, P.~Krah and R.~Sommer, \emph{{SU(3) Yang Mills theory at
  small distances and fine lattices}},
  \href{https://doi.org/10.1051/epjconf/201817514024}{\emph{EPJ Web Conf.}
  {\bfseries 175} (2018) 14024}
  [\href{https://arxiv.org/abs/1711.01860}{{\ttfamily 1711.01860}}].
\bibitem{Brambilla:2023fsi}
N.~Brambilla, V.~Leino, J.~Mayer-Steudte and A.~Vairo, \emph{{Static force from
  generalized Wilson loops on the lattice using the gradient flow}},
  \href{https://doi.org/10.1103/PhysRevD.109.114517}{\emph{Phys. Rev. D}
  {\bfseries 109} (2024) 114517}
  [\href{https://arxiv.org/abs/2312.17231}{{\ttfamily 2312.17231}}].
\bibitem{Jansen:2011vv}
{\scshape ETM} collaboration, K.~Jansen, F.~Karbstein, A.~Nagy and M.~Wagner,
  \emph{{$\Lambda_{\overline{\rm MS}}$ from the static potential for QCD with
  $n_f=2$ dynamical quark flavors}},
  \href{https://doi.org/10.1007/JHEP01(2012)025}{\emph{JHEP} {\bfseries 01}
  (2012) 025} [\href{https://arxiv.org/abs/1110.6859}{{\ttfamily 1110.6859}}].
\bibitem{Karbstein:2014bsa}
F.~Karbstein, A.~Peters and M.~Wagner,
  \emph{{${\Lambda}_{\overline{\mathrm{MS}}}^{({n}_f=2)}$ from a momentum space
  analysis of the quark-antiquark static potential}},
  \href{https://doi.org/10.1007/JHEP09(2014)114}{\emph{JHEP} {\bfseries 09}
  (2014) 114} [\href{https://arxiv.org/abs/1407.7503}{{\ttfamily 1407.7503}}].
\bibitem{Karbstein:2018mzo}
F.~Karbstein, M.~Wagner and M.~Weber, \emph{{Determination of
  $\Lambda_{\overline{\textrm{MS}}}^{(n_f=2)}$ and analytic parametrization of
  the static quark-antiquark potential}},
  \href{https://doi.org/10.1103/PhysRevD.98.114506}{\emph{Phys. Rev. D}
  {\bfseries 98} (2018) 114506}
  [\href{https://arxiv.org/abs/1804.10909}{{\ttfamily 1804.10909}}].
\bibitem{Bazavov:2012ka}
A.~Bazavov, N.~Brambilla, X.~Garcia~i Tormo, P.~Petreczky, J.~Soto and
  A.~Vairo, \emph{{Determination of $\alpha_s$ from the QCD static energy}},
  \href{https://doi.org/10.1103/PhysRevD.86.114031}{\emph{Phys. Rev. D}
  {\bfseries 86} (2012) 114031}
  [\href{https://arxiv.org/abs/1205.6155}{{\ttfamily 1205.6155}}].
\bibitem{Bazavov:2014soa}
A.~Bazavov, N.~Brambilla, X.~Garcia~i Tormo, P.~Petreczky, J.~Soto and
  A.~Vairo, \emph{Determination of {$\alpha_{\text{s}}$} from the {QCD} static
  energy: An update},
  \href{https://doi.org/10.1103/PhysRevD.90.074038}{\emph{Phys. Rev. D}
  {\bfseries 90} (2014) 074038}
  [\href{https://arxiv.org/abs/1407.8437}{{\ttfamily 1407.8437}}].
\bibitem{Takaura:2018vcy}
H.~Takaura, T.~Kaneko, Y.~Kiyo and Y.~Sumino, \emph{Determination of
  {$\alpha_{\text{s}}$} from static {QCD} potential: {OPE} with renormalon
  subtraction and lattice {QCD}},
  \href{https://doi.org/10.1007/JHEP04(2019)155}{\emph{JHEP} {\bfseries 04}
  (2019) 155} [\href{https://arxiv.org/abs/1808.01643}{{\ttfamily
  1808.01643}}].
\bibitem{Bazavov:2019qoo}
{\scshape TUMQCD} collaboration, A.~Bazavov, N.~Brambilla, X.~Garcia~i Tormo,
  P.~Petreczky, J.~Soto, A.~Vairo et~al., \emph{Determination of the {QCD}
  coupling from the static energy and the free energy},
  \href{https://doi.org/10.1103/PhysRevD.100.114511}{\emph{Phys. Rev. D}
  {\bfseries 100} (2019) 114511}
  [\href{https://arxiv.org/abs/1907.11747}{{\ttfamily 1907.11747}}].
\bibitem{Ayala:2020odx}
C.~Ayala, X.~Lobregat and A.~Pineda, \emph{Determination of {$\alpha(M_{Z})$}
  from an hyperasymptotic approximation to the energy of a static
  quark-antiquark pair},
  \href{https://doi.org/10.1007/JHEP09(2020)016}{\emph{JHEP} {\bfseries 09}
  (2020) 016} [\href{https://arxiv.org/abs/2005.12301}{{\ttfamily
  2005.12301}}].
\bibitem{Mena-Valle:2025hky}
J.~M. Mena-Valle, V.~Mateu and P.~G. Ortega, \emph{{A Precise $\alpha_s$
  Determination from the R-improved QCD Static Energy}},
  \href{https://arxiv.org/abs/2510.24846}{{\ttfamily 2510.24846}}.
\bibitem{Leino:2025pvl}
{\scshape TUMQCD} collaboration, V.~Leino, A.~Bazavov, N.~Brambilla, A.~S.
  Kronfeld, J.~Mayer-Steudte, P.~Petreczky et~al., \emph{{Strong coupling in
  (2+1+1)-flavor QCD}}, \href{https://doi.org/10.22323/1.466.0298}{\emph{PoS}
  {\bfseries LATTICE2024} (2025) 298}
  [\href{https://arxiv.org/abs/2502.01453}{{\ttfamily 2502.01453}}].
\bibitem{Brambilla:2021wqs}
N.~Brambilla, V.~Leino, O.~Philipsen, C.~Reisinger, A.~Vairo and M.~Wagner,
  \emph{{Lattice gauge theory computation of the static force}},
  \href{https://doi.org/10.1103/PhysRevD.105.054514}{\emph{Phys. Rev. D}
  {\bfseries 105} (2022) 054514}
  [\href{https://arxiv.org/abs/2106.01794}{{\ttfamily 2106.01794}}].
\bibitem{Brambilla:2017hcq}
{\scshape TUMQCD} collaboration, N.~Brambilla, J.~Komijani, A.~S. Kronfeld and
  A.~Vairo, \emph{Relations between heavy-light meson and quark masses},
  \href{https://doi.org/10.1103/PhysRevD.97.034503}{\emph{Phys. Rev. D}
  {\bfseries 97} (2018) 034503}
  [\href{https://arxiv.org/abs/1712.04983}{{\ttfamily 1712.04983}}].
\bibitem{Komijani:2017vep}
J.~Komijani, \emph{{A discussion on leading renormalon in the pole mass}},
  \href{https://doi.org/10.1007/JHEP08(2017)062}{\emph{JHEP} {\bfseries 08}
  (2017) 062} [\href{https://arxiv.org/abs/1701.00347}{{\ttfamily
  1701.00347}}].
\bibitem{Kronfeld:2023jab}
A.~S. Kronfeld, \emph{{Factorial growth at low orders in perturbative QCD:
  control over truncation uncertainties}},
  \href{https://doi.org/10.1007/JHEP12(2023)108}{\emph{JHEP} {\bfseries 12}
  (2023) 108} [\href{https://arxiv.org/abs/2310.15137}{{\ttfamily
  2310.15137}}].
\bibitem{Curci:1983an}
G.~Curci, P.~Menotti and G.~Paffuti, \emph{{Symanzik's Improved Lagrangian for
  Lattice Gauge Theory}},
  \href{https://doi.org/10.1016/0370-2693(83)91043-2}{\emph{Phys. Lett. B}
  {\bfseries 130} (1983) 205}.
\bibitem{Heller:1984hx}
U.~M. Heller and F.~Karsch, \emph{{One Loop Perturbative Calculation of Wilson
  Loops on Finite Lattices}},
  \href{https://doi.org/10.1016/0550-3213(85)90261-5}{\emph{Nucl. Phys. B}
  {\bfseries 251} (1985) 254}.
\bibitem{Weisz:1983bn}
P.~Weisz and R.~Wohlert, \emph{{Continuum Limit Improved Lattice Action for
  Pure Yang-Mills Theory. 2.}},
  \href{https://doi.org/10.1016/0550-3213(84)90563-7}{\emph{Nucl. Phys. B}
  {\bfseries 236} (1984) 397}.
\bibitem{Snippe:1997ru}
J.~R. Snippe, \emph{{Computation of the one loop Symanzik coefficients for the
  square action}},
  \href{https://doi.org/10.1016/S0550-3213(97)00270-8}{\emph{Nucl. Phys. B}
  {\bfseries 498} (1997) 347}
  [\href{https://arxiv.org/abs/hep-lat/9701002}{{\ttfamily hep-lat/9701002}}].
\bibitem{Martinelli:1998vt}
G.~Martinelli and C.~T. Sachrajda, \emph{{Computation of the b quark mass with
  perturbative matching at the next-to-next-to-leading order}},
  \href{https://doi.org/10.1016/S0550-3213(99)00423-X}{\emph{Nucl. Phys. B}
  {\bfseries 559} (1999) 429}
  [\href{https://arxiv.org/abs/hep-lat/9812001}{{\ttfamily hep-lat/9812001}}].
\bibitem{Athenodorou:2005hi}
A.~Athenodorou and H.~Panagopoulos, \emph{{Large Wilson loops with overlap and
  clover fermions: Two-loop evaluation of the b-quark mass shift and the
  quark-antiquark potential}},
  \href{https://doi.org/10.1016/j.nuclphysb.2008.02.006}{\emph{Nucl. Phys. B}
  {\bfseries 799} (2008) 1}
  [\href{https://arxiv.org/abs/hep-lat/0509039}{{\ttfamily hep-lat/0509039}}].
\bibitem{Bali:2002wf}
G.~S. Bali and P.~Boyle, \emph{Perturbative {Wilson} loops with massive sea
  quarks on the lattice},
  \href{https://arxiv.org/abs/hep-lat/0210033}{{\ttfamily hep-lat/0210033}}.
\bibitem{Follana:2006rc}
{\scshape HPQCD} collaboration, E.~Follana, Q.~Mason, C.~Davies, K.~Hornbostel,
  G.~P. Lepage, J.~Shigemitsu et~al., \emph{Highly improved staggered quarks on
  the lattice, with applications to charm physics},
  \href{https://doi.org/10.1103/PhysRevD.75.054502}{\emph{Phys. Rev. D}
  {\bfseries 75} (2007) 054502}
  [\href{https://arxiv.org/abs/hep-lat/0610092}{{\ttfamily hep-lat/0610092}}].
\bibitem{Hart:2009nr}
A.~Hart, G.~M. von Hippel, R.~R. Horgan and E.~H. Muller, \emph{Automated
  generation of lattice {QCD} feynman rules},
  \href{https://doi.org/10.1016/j.cpc.2009.04.021}{\emph{Comput. Phys. Commun.}
  {\bfseries 180} (2009) 2698}
  [\href{https://arxiv.org/abs/0904.0375}{{\ttfamily 0904.0375}}].
\bibitem{Eiras:1999xx}
D.~Eiras and J.~Soto, \emph{Effective field theory approach to pionium},
  \href{https://doi.org/10.1103/PhysRevD.61.114027}{\emph{Phys. Rev. D}
  {\bfseries 61} (2000) 114027}
  [\href{https://arxiv.org/abs/hep-ph/9905543}{{\ttfamily hep-ph/9905543}}].
\bibitem{lpt_paper}
G.~M. von Hippel, V.~Leino and S.~Steinbeißer, \emph{{One loop improvement of
  the static potential with HISQ quarks}}, {\emph{In preparation: TUM-EFT
  171/22} (2026) }.
\bibitem{Larsen:2025wvg}
R.~Larsen, S.~Mukherjee, P.~Petreczky, H.-T. Shu and J.~H. Weber, \emph{{Scale
  Setting and Strong Coupling Determination in the Gradient Flow Scheme for 2+1
  Flavor Lattice QCD}},  \href{https://arxiv.org/abs/2502.08061}{{\ttfamily
  2502.08061}}.
\bibitem{Jay:2020jkz}
W.~I. Jay and E.~T. Neil, \emph{{Bayesian} model averaging for analysis of
  lattice field theory results},
  \href{https://doi.org/10.1103/PhysRevD.103.114502}{\emph{Phys. Rev. D}
  {\bfseries 103} (2021) 114502}
  [\href{https://arxiv.org/abs/2008.01069}{{\ttfamily 2008.01069}}].
\end{thebibliography}\endgroup

\end{document}